\documentstyle{l-aa}
\input{epsf.sty}

\begin{document}

\thesaurus{10.15.2: NGC 4815;08.06.03;08.12.3;10.19.2;03.13.6}

\title{
The spatial distribution and luminosity function of the  open
cluster NGC 4815
}

\author{B. Chen \inst{1} \and G. Carraro \inst{2,3} \and J. Torra
\inst{1} \and C. Jordi \inst{1} }

\offprints{bchen@mizar.am.ub.es}

\institute{Departament d'Astronomia i Meteorologia,
           Universitat de Barcelona,
           Avda. Diagonal 647,
           E08028, Barcelona, Spain
\and  Department of Astronomy, Padova University, Vicolo dell'Osservatorio
5, I-35122 Padova, Italy
\and SISSA/ISAS, via Beirut 2, I-34013, Trieste, Italy
}
\date{Received August 22; Accepted November 28, 1997 }
\maketitle
\begin{abstract}
NGC 4815 is a distant and populous open cluster, which lies in the
galactic plane in a region of strong 
absorption. As a consequence, its membership, spatial distribution and
luminosity 
function are not well determined. In this paper, 
we present an  algorithm which uses both positional 
and photometric data simultaneously to search for  open cluster members.
The contribution of the field stars is estimated by our Galaxy model software.
The method has been developed for deep CCD photometric observations,
but it can also be         
used to investigate the membership  in an n-dimensional 
space, including proper motions, radial velocity and
metallicity.

A clean and well defined colour-magnitude diagram (CMD) for NGC 4815  
has been obtained after decontaminating from
field stars. 
We  have investigated the spatial  distribution of cluster members
and the luminosity function. We have shown that the projected 
surface distribution of 
stars is well-represented by an exponentially decreasing function with a 
scale length of 1.67 $\pm$ 0.06 arcmin. 
We have found evidence for mass segregation. 
The luminosity function for NGC 4815 is 
similar to that of the Hyades cluster, and shows a gap at $V$ = 15 mag.

\end{abstract}

\section{Introduction}

Galactic open clusters can be used to test both 
the theories of stellar evolution (Carraro et al. 1994) and the 
formation and kinematics of the Galactic 
disk (Palous et al. 1977; Sandage 1988). 
The advent of CCDs, with high sensitivity and 
linearity, has allowed observers to obtain deep photometry of open clusters.  
This is particularly important for distant clusters or clusters
located at low galactic
latitudes and highly contaminated by foreground stars
(Carraro \& Patat 1995).
This is the case for
NGC 4815 ($l$ = 303$^o$, $b$=-2$^o$), a distant open cluster lying in the
galactic plane,
in a region of strong absorption.

The first photometric observations of NGC 4815 were carried out
by Moffat \& Vogt (1973), who measured only 9 stars and did 
not recognize any sequence.
Kjeldsen \& Frandsen (1991) have obtained $UBV$ - CCD photometry of NGC 4815,
including 599 stars, but, because of the contamination of noncluster
members, they could not determine cluster membership, reddening or age.
More recently, Carraro \& Ortolani (1994, hereafter CO94) have obtained 
deeper $BV$ - CCD photometry with the NTT 3.5 m telescope at the European 
Southern Observatory (ESO).  Their sample includes 2498 stars.

Photometric observations have also been published 
by Phelps et al. (1994) in the
context of a survey of old open clusters. Unfortunately their 
observations were made with a small 
telescope (0.9 m) and 
the errors in the photometric data do not allow an accurate determination 
of the physical properties of the cluster.
To our knowledge, no proper motion study has been 
published in the literature. 
So, the studies of
Kjeldsen \& Frandsen (1991) and CO94
will be adopted as the reference photometric studies for this paper.

In order to investigate the cluster properties of NGC 4815,  
CO94 considered all stars within
$4^{\prime}.6$ of the center to be 
open cluster members. 
The CMD of their cluster members (see Fig. 4 of CO94)
is very similar to that of the total sample
(see Fig. 2 of CO94).
The red stars ($B-V$ $>$ 1.5 and $V$ $>$16) in their CMD of cluster 
members are in fact non-cluster 
members from the stellar evolutionary point of view. Therefore, they 
cannot reliably discuss the spatial distribution and luminosity function
of NGC 4815 
because of the many field stars in the sample, and their conclusions must be 
considered as preliminary.

Large-format CCD observations can provide deep star counts and 
extend to the edge of the cluster, but the larger field and fainter magnitude limit
also means that cluster sequences are contaminated with many field stars.
For old open clusters, it is well-known that the brighter stars are more 
centrally concentrated (Reid, 1992). Thus,
the radial distance from the cluster centre is not a suitable criterion 
by which  to classify members: by 
choosing
a large radius, the contamination of field stars increases, but by using a
small radius, 
one can undersample the low mass stars which are not    
centrally  concentrated, and thus bias the luminosity function.
Therefore the development of an  algorithm, which uses
all information available (positions, photometry, and/or proper motions)
simultaneously to search for  
open cluster members, 
is a necessary task prior to any study of many problems in open clusters.

In this paper, we present  a new 
algorithm from pattern recognition theory. The advantage of our algorithm 
is twofold. First, we use all available 
information:
apparent magnitude ($V$), colour ($B-V$), and positions ($X$,$Y$) to
extract open cluster members in a four-dimensional space.  
Second, we 
separate open cluster members from field stars without assuming 
any {\sl  a priori} distribution of the open cluster members. 
This is important for our studies about the spatial distribution 
and luminosity function of the cluster. 
We should point out that the method described in this paper is well suited to 
identifying members in a statistical sense, and hence allows  one to study the 
statistical properties of the cluster. Additional methods clearly need to be 
used if an analysis requires  only cluster members.

The paper is organized as follows: 
in Sect. 2, 
we briefly describe the structure parameters in our Galaxy model.  
In Sect. 3, we develop a new cluster member finding algorithm, 
and investigate  the membership of open cluster NGC 4815. 
We have compared our results with those  of CO94
and Kjeldsen \& Frandsen (1991).
In Sect. 4, the spatial
distribution of cluster members and the luminosity function are 
discussed. 
Finally, we
summarize the main conclusions in Sect. 5.

\section{The Galaxy Model}

The difficulty in cluster membership identification is the
contamination by the field stars. In this paper, 
we use our Galaxy model  to predict the
distribution of the field stars in the direction of the cluster, 
and establish a classification system for identifying
cluster members. 

\subsection{The starcounts Model}

The main characteristics of stellar populations have been extracted from many 
observations, and  
the application of computer modeling to observations 
(Bahcall \& Soneira 1980; Pritchet 1983; Gilmore 1984; 
Robin \& Cr\'ez\'e 1986; Ratnatunga et al. 1989; Reid \& Majewski 1993; 
Chen 1997a,b) has generated considerable interest.
Models can  parameterize all stellar populations and allow us to
calculate the expected star counts explicitly.

We have constructed a  Galactic structure and kinematic model 
(Chen 1997a,b),
which includes a thin disk,
a thick disk, and a halo.
The model can 
predict the magnitudes, positions, colours, proper motions,
radial velocities,
metallicities according to the selection 
criteria used in the  observation. 
A detailed description, including kinematical parameters (proper motions, 
radial velocity, space velocity), can be found in Chen (1997b). 
In the following lines we give a short description of the main structure 
parameters.

The method used in our Galaxy model is the numerical integration of the 
fundamental equation of stellar statistics, for the j-th subsystem; 

\begin{equation}
A_{j}(V, B-V) = \omega \int \Psi_{j} (M) D_{j} (r) r^{2} dr
\end{equation}

Where $A_{j} (V, B-V)$ is the number of stars of a given colour index at
a given 
apparent magnitude; $\Psi_{j}$ ($M$) is the luminosity 
function and $D_{j}$($r$), 
the density law. In order  to
predict star counts in a 
given direction, a luminosity function and density distribution 
for each population or sub-population in the model is required. 
The  stellar density laws used for the thin and
thick 
disk are exponential. For the population II spheroid stars, the density law is 
adopted following de Vaucouleurs (1977). We adopt the luminosity function 
for the thin  
and thick disk stars given by Wielen et al. (1983).  The halo luminosity 
function is assumed to be similar in shape to that of the 
disk for $M_{v}$ $\ge$ 4, but matching a globular cluster 
function at brighter magnitudes (Reid, 1993a).     
We adopt a CMD given by Bahcall et al. (1987) from the Yale 
Parallax catalog for thin and thick disk stars.   For halo stars, 
the colour-magnitude diagram for both the main-sequence and evolved (red giant 
and horizontal-branch) stars is based on the table given by 
Bergbush \& Vandenberg (1992).  

The thin and thick disk giants were included separately in the model. 
According to Bahcall \& Soneira (1981), 
we take the fraction of stars, $f$, on the main sequence (MS) in the plane 
of the disk to be:

$f=0.44 e^{0.00015 (M_{v}+8) ^{3.5}}$,   $M_{v}$ $<$ 3.7

$f=1$,                              $M_{v}$ $\ge$ 3.7

We have allowed  for observational uncertainties by adding Gaussian errors
to the $V$ magnitude  and to the
colour of each 'star' generated.
CO94 have obtained errors of 0.02, 0.05, and 0.07
mag at $V$=15.5, 17.5 and 19.5 mag respectively, and at the same magnitude levels, 
they found dispersions in colour of 0.05, 0.11 and 0.14 mag, respectively.
These results were incorporated in our Galaxy model software.

We should point out that these parameters in the model code  
are not the results of  this 
investigation, but have been derived  from previously
published results (Bahcall \& Soneira 1980; Robin \& Cr\'ez\'e 1986;
Reid \& Majewski 1993; Chen 1997a; Chen 1997b).

\subsection{Extinction in the line of sight}

One of the important characteristics of galactic obscuration is that 
it is patchy, so,  it is important to discuss the foreground obscuration 
in the region of NGC 4815 ($d$ $\sim$ 2300 pc).  
Neckel \& Klare (1980) have derived extinctions and distances
for more than 11000 stars and  investigated the spatial
distribution of the interstellar extinction at $\mid$ $b$ $\mid$  $<$ 7$^o$.6.
Their results are shown in a series of diagrams.
In Fig. 1, we plot their results (open circles) from the nearest field
($l$ = 302$^o$, $b$ = -3$^o$) in the direction of NGC 4815.
Using a large sample (about 17000 stars)
with  MK spectral types and photoelectric photometry, Arenou et al. (1992)
constructed a tridimensional model of the galactic interstellar 
extinction.  In Fig. 1, we show the results (solid line) derived
by Arenou et al. (1992).

Recently, we  
have used  Lund Catalogue of 
Open Cluster Data (Lyng\a{a} 1987) to derive an analytic expression of the 
interstellar extinction (Chen et al. 1997b).  
In Fig. 1, we have overplotted the average extinction in the galactic
plane (filled circles).
We can see that extinction laws derived from different 
methods are fairly consistent. 
Moreover, Bertelli et al. (1995)  derived the extinction 
in the line of sight in the
field near the cluster Lyng\a{a} 7 ($l$ = 328$^o$, $b$ = -2$^o$.8), 
not far from the 
field
of our investigation. They found an  extinction of $A_{v}$ = 1.5 mag at 
2000 pc, and $A_{v}$ = 2 mag at 4000 pc,
which is also consistent with the results shown in Fig. 1.

In Fig.~2, we show the observed CMD for all the stars 
in a field of view of 12'.3 x 12'.3 from CO94. 
We can recognize the MS and the  giant branch. 
Many field stars seem to contaminate the 
CMD, especially a red sequence parallel to the MS.

In Fig. 3, we show the  simulated colour-magnitude diagram 
from the Galaxy model with the Arenou et al. (1992)  
extinction law.
This predicted diagram provides us with the distribution of 
field stars in this 
direction.  
In following discussion,
we adopt  the extinction law derived by Arenou et al. (1992) in our Galaxy 
model.

   \begin{figure}
    \begin{center}
    \leavevmode
    \epsfxsize=18cm
    \epsfbox{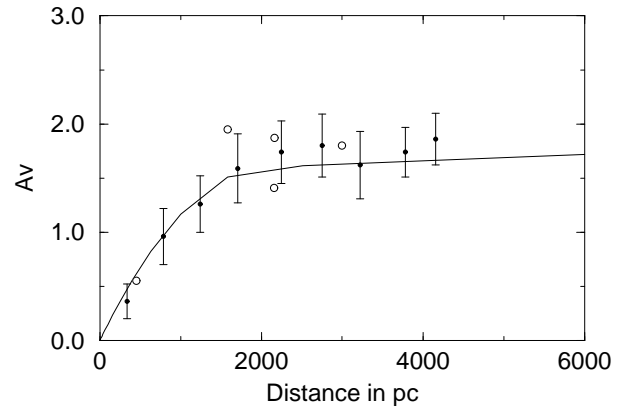}
    \end{center}
     \caption[]{The extinction along the line of sight in 
     NGC 4815. Solid line is the Arenou et al (1992) extinction law, 
filled circles are the 
results derived by us (Chen et al. 1997b) from open cluster system,
open circles are the results of Neckel \& Klare (1980). NGC 4815 is located 
at about 2300 pc from the Sun.}
   \end{figure}

   \begin{figure}
    \begin{center}
    \leavevmode
    \epsfxsize=18cm
    \epsfbox{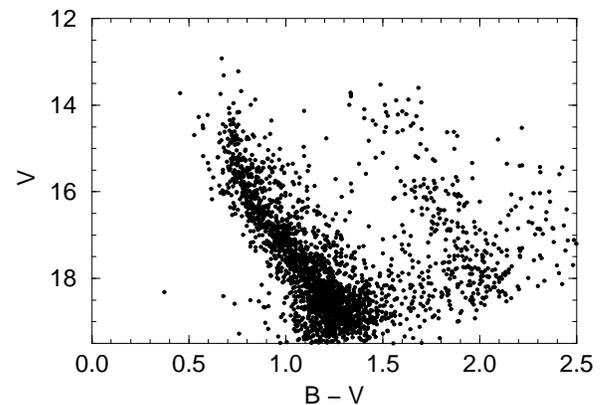}
    \end{center}
     \caption[]{Observed CMD for all the stars studied in the region 
      of NGC 4815} 
   \end{figure}

   \begin{figure}
    \begin{center}
    \leavevmode
    \epsfxsize=18cm
    \epsfbox{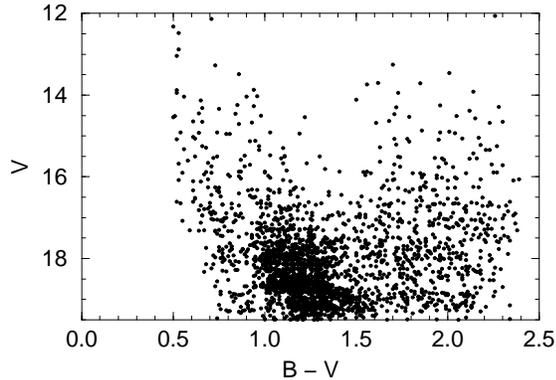}
    \end{center}
     \caption[]{ Simulated CMD in the region of NGC 4815}
   \end{figure}

\section{Identification of NGC 4815 members}

Many authors  have used colour-magnitude 
and colour-colour 
diagrams to separate open cluster members and field stars. 
When kinematic data are available, it is commonly accepted that the 
membership probabilities obtained from the analysis of 
proper motions or radial velocities are more reliable. The 
parametric models proposed by 
Vasilevskis et al. (1958) and Sanders (1971) assume that the 
proper motion distributions of field stars and cluster members 
can be modeled by a normal bivariate function. However, in several 
cases the wings of the field star distribution are more extensively populated  
than expected based on the predictions of a normal bivariate function 
(Marschall \& van Altena, 1987).    
Cabrera-Ca\~no \& Alfaro (1990) have developed a non-parametric 
approach to identify  open cluster members by  using  proper motions 
and positions as  variables.  
However,  because their method assumes that the variables are independent,
it cannot be used with photometric data.   
\\
\\
In this paper, we have developed a new 
cluster member finding algorithm. This method, based on a  non-parametric 
approach, has been previously used to    
identify moving group members in the solar neighborhood 
(Chen et al. 1997a) and to isolate  galactic thick disk stars from star count 
survey 
(Chen et al. 1992; Chen 1996a).  
The viability of the method has been tested by a series of Monte Carlo 
simulations.  Results show that the method is powerful.
The detailed descriptions and numerical simulations of the method can be 
found in 
Chen (1996b) and Chen et al. (1997a).

\subsection{The method}

Let ({\vec x$_{1}$}, {\vec x$_{2}$}, {\vec x$_{3}$}, ... {\vec x$_{n}$})
be a set of  stars
to be studied. Each star {\vec x$_{i}$} is considered  as one point
in the four-dimensional space  $(X, Y, V, B-V)$ of position,
magnitude and colour.

Let $N_{1}$ and
$N_{2}$ be the number
of stars in the observed and the 'simulated' sample, respectively.
The number of stars ($N_{2}$)  comes from the
Galaxy model and is considered as the contribution to the counts of the
field stars in the direction of the open cluster.
According to the Bayesian theory,
the probability of a given star being an open cluster member can be
derived as:

\begin{equation}
P(c|{\vec x_{i}}) = 1 - P(f|{\vec x_{i}}) = 1 - \frac{N_{2}
p({\vec x_{i}}|f)}{N_{1} p({\vec x_{i}})}
\end{equation}

Where $P(c|{\vec x_i})$ and $P(f | {\vec x_i})$ are the
 {\sl a posteriori} probability that an object
{\vec x$_{i}$} belongs to the cluster and field stars, respectively.
$p$({\vec x$_{i}$}) is the probability density function (hereafter pdf)
from the observed sample, $p$({\vec x$_{i}$}$\mid$$f$) is the class
conditional pdf for
field stars. The probability
$P$(c$\mid${\vec x$_{i}$})
can be derived by calculating $p$({\vec x$_{i}$})
and $p$({\vec x$_{i}$}$\mid$f) at each
observed star {\vec x$_{i}$}.

The  true $p$({\vec  x$_{i}$}) in each star
can be estimated by a kernel
estimator $\hat{p}$ ({\vec x$_{i}$}) of the
density (Hand, 1982):

\begin{equation}
\hat{p}({\vec x_i})
= \frac{1}{nh^d}\sum_{j=1}^n
\frac{1}{{| \Sigma |}^{1/2}
({2\pi})^{d/2}}e^{-\frac{1}{2h^2}
({\vec x_i}-{\vec x_j})^{'}{ \Sigma}^{-1}({\vec x_i}-{\vec x_j})}
\end{equation}

    \noindent where  {\vec x$_{i}$} is  the point at which the
estimate is being made,
$n$ is the number of stars in the observed sample,
{\vec x$_{j}$} is the observed sample set defined in
d-dimensional space,
({\vec x$_{i}$} -{\vec x$_{j}$})' is the transpose of the vector
({\vec x$_{i}$} -{\vec x$_{j}$}) and
$\Sigma$ is the variance-covariance matrix of the observed sample.

The optimal smoothing parameter $h$ derived
by Silverman (1986) can be written as:

\begin{equation}
h=(\frac{4}{d+2})^{1/(d+4)} \sigma n^{-1/(d+4)}
\end{equation}

\noindent    where $\sigma$ is the average marginal variance,  $\sigma^{2}
= d^{-1}\Sigma \sigma_{i}^{2}$.

    From  the  definition  of  the  kernel  estimator, we know that
    it is a sum of
     the  multivariate normal density functions
    placed  at
    the observations.  The functions  determine
    the shape while the  smoothing parameter $h$ determines  their
    width.
Concerning  the computation of the class conditional probability
density function, $p({\vec x_{i}} \mid f)$, the procedure is the
same as described above for $p({\vec x_{i}})$, but
we have to perform the summation over the simulated sample for the
field stars.

   \begin{figure*}
    \begin{center}
    \leavevmode
    \epsfxsize=16cm
    \epsfbox{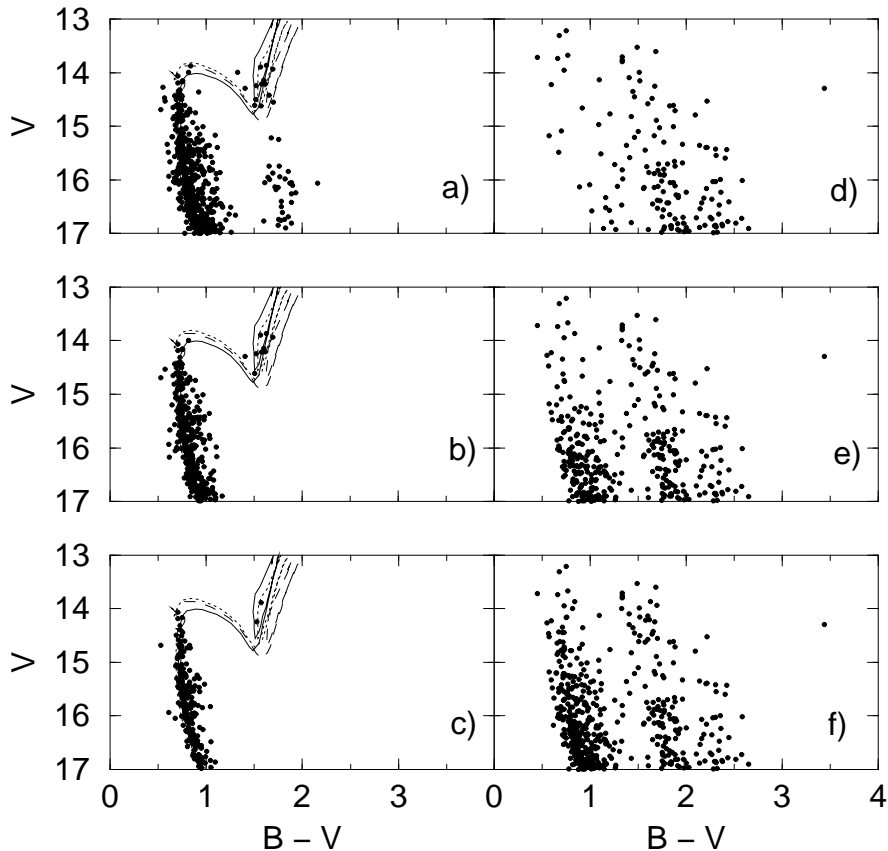}
    \end{center}
     \caption[]{ {\sl Left panel:}
      The CMD for the cluster
        members, the Arenou et al. (1992) extinction law has been adopted.
     Superimposed are
     isochrones of different metal abundance $Z$. Dashed line represents a
     $Z$= 0.008 isochrone for an age of 500 Myr, plotted adopting
     $E_{B-V}$ = 0.70 and ($V-M_{v}$) = 14.2. Solid line represents a
     $Z$ = 0.004 isochrone for an age of 600 Myr, plotted adopting
     $E_{B-V}$ = 0.75 and ($V-M_{v}$) = 14.10, dotted line represents
     a $Z$ = 0.02 isochrone for an age of 500 Myr, plotted adopting
    $E_{B-V}$ = 0.65 and ($V-M_{v}$) = 14.20. a) The cluster members with
       probability $P$(c$\mid$${\vec x_i}$) $>$ 0.5, b) The cluster members 
    with  probability $P$(c$\mid$${\vec x_i}$) $>$ 0.8, 
    c) The cluster members with
        probability $P$(c$\mid$${\vec x_i}$) $>$ 0.9 \\
     {\sl Right panel:} The CMD for the field stars. d) The stars
    with probability $P$(c$\mid$${\vec x_i}$) $\le$ 0.5, e) The stars with
       probability $P$(c$\mid$${\vec x_i}$) $\le$ 0.8, f) The stars with
        probability $P$(c$\mid$${\vec x_i}$) $\le$ 0.9}

   \end{figure*}

   \begin{figure*}
    \begin{center}
    \leavevmode
    \epsfxsize=16cm
    \epsfbox{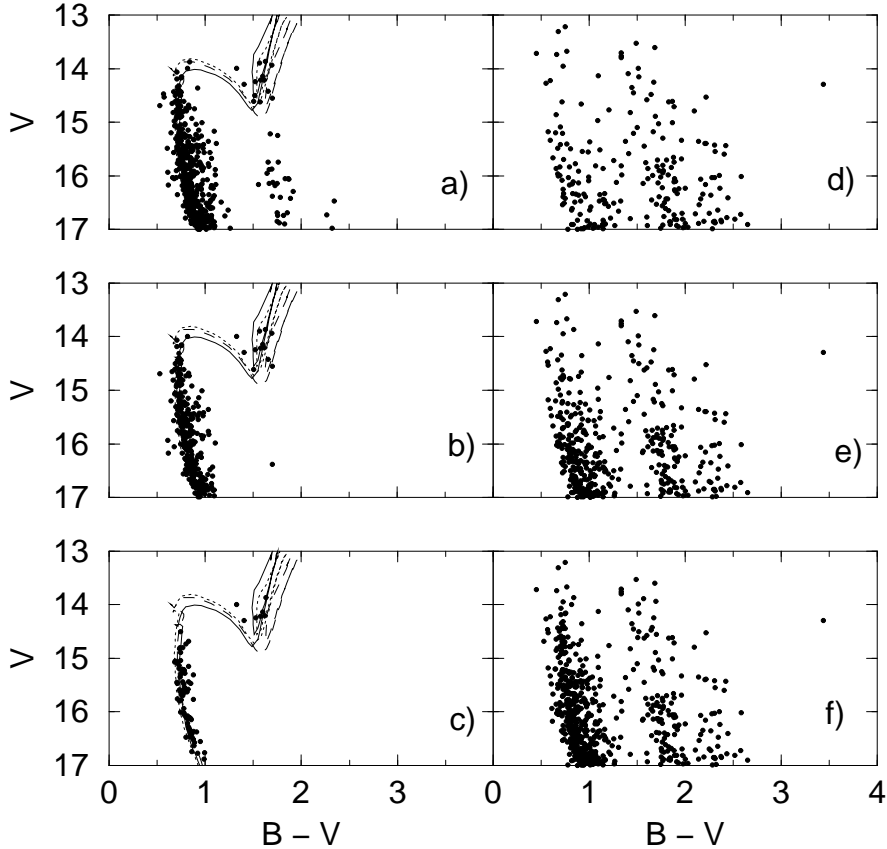}
    \end{center}
     \caption[]{ as Fig 4, but using the Arp (1965) reddening distribution.}
   \end{figure*}

   \begin{figure}
    \begin{center}
    \leavevmode
    \epsfxsize=18cm
    \epsfbox{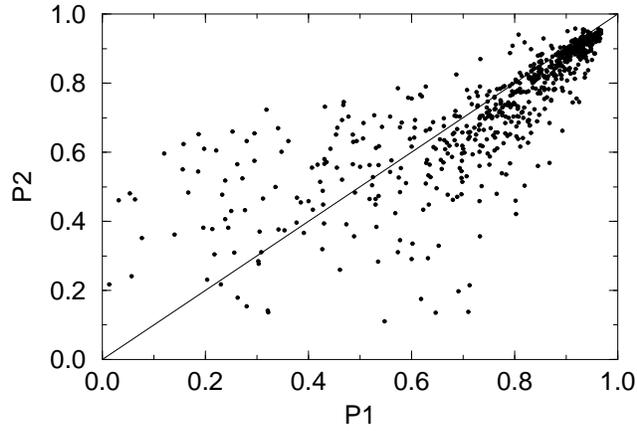}
    \end{center}
     \caption[]{ The probability P1 derived from 
the Arenou et al. (1992) extinction model against          
the probability P2 derived from the Arp (1965) extinction model}  
   \end{figure}

   \begin{figure}
    \begin{center}
    \leavevmode
    \epsfxsize=18cm
    \epsfbox{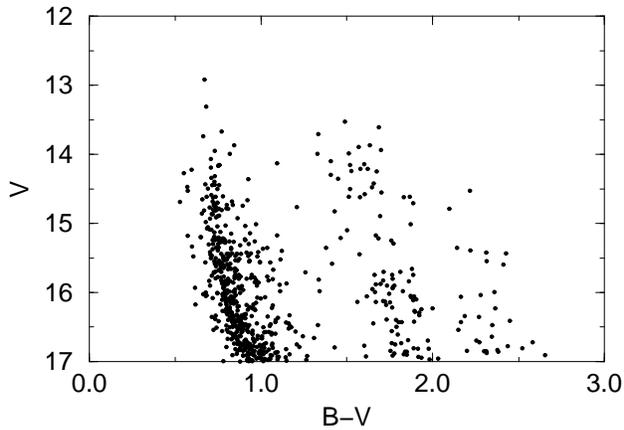}
    \end{center}
     \caption[]{ CMD of NGC 4815 for cluster members derived
from the criteria of CO94, i.e. stars within a radius
of 4.6 arcmin of cluster center.}
   \end{figure}

   \begin{figure}
    \begin{center}
    \leavevmode
    \epsfxsize=18cm
    \epsfbox{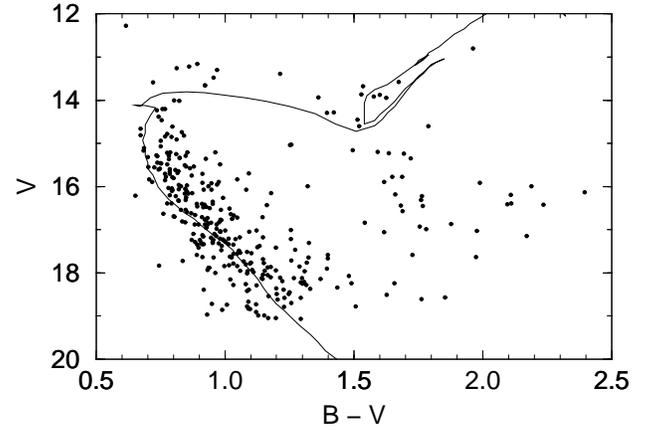}
    \end{center}
     \caption[]{ CMD obtained by Kjeldsen \& 
   Frandsen (1991). Superimposed is isochrone of $Z$ = 0.008      
   for age = 500 Myr, plotted adopting $E_{B-V}$= 0.7 and 
    $V-M_{v}$=14.2 mag}
   \end{figure}

   \begin{figure}
    \begin{center}
    \leavevmode
    \epsfxsize=18cm
    \epsfbox{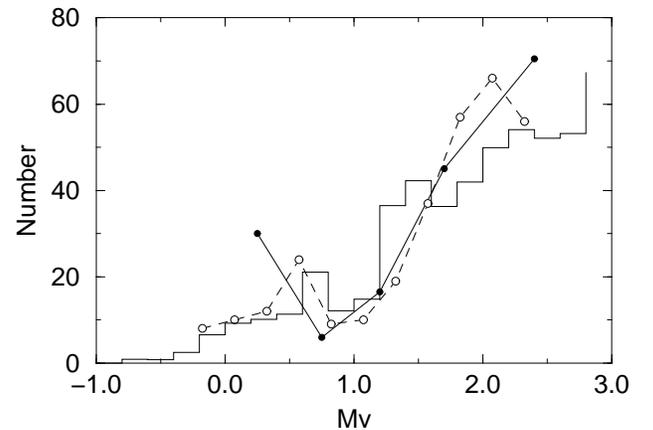}
    \end{center}
     \caption[]{ The luminosity function of the cluster NGC 4815.
    The luminosity functions of the Hyades  (filled circles)  and
    NGC 7789 (open circles), which are shifted arbitrarily to
            match the bright part,
         have been overplotted}
   \end{figure}

   \begin{figure}
    \begin{center}
    \leavevmode
    \epsfxsize=18cm
    \epsfbox{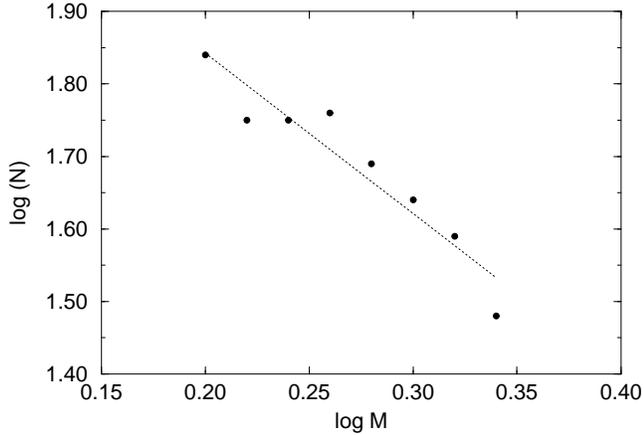}
    \end{center}
     \caption[]{ The Mass  function of the cluster NGC 4815. The 
     solid line is the result from least-squares fit}
   \end{figure}

   \begin{figure}
    \begin{center}
    \leavevmode
    \epsfxsize=18cm
    \epsfbox{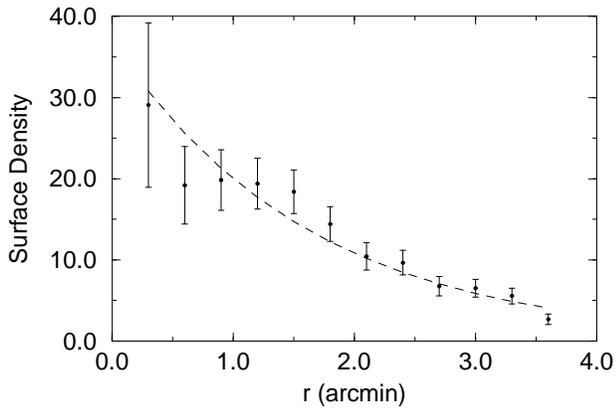}
    \end{center}
     \caption[]{ The radial density distribution for the cluster 
   NGC 4815, where the error bars were calculated assuming that 
   the number of stars in a bin is governed by Poisson statistics, 
    the dotted line we have adopted to  fit the spatial distribution is an 
    exponentially decreasing function of the distance from the 
     cluster center with a scale length of 1.67 arcmin} 
   \end{figure}

   \begin{figure}
    \begin{center}
    \leavevmode
    \epsfxsize=18cm
    \epsfbox{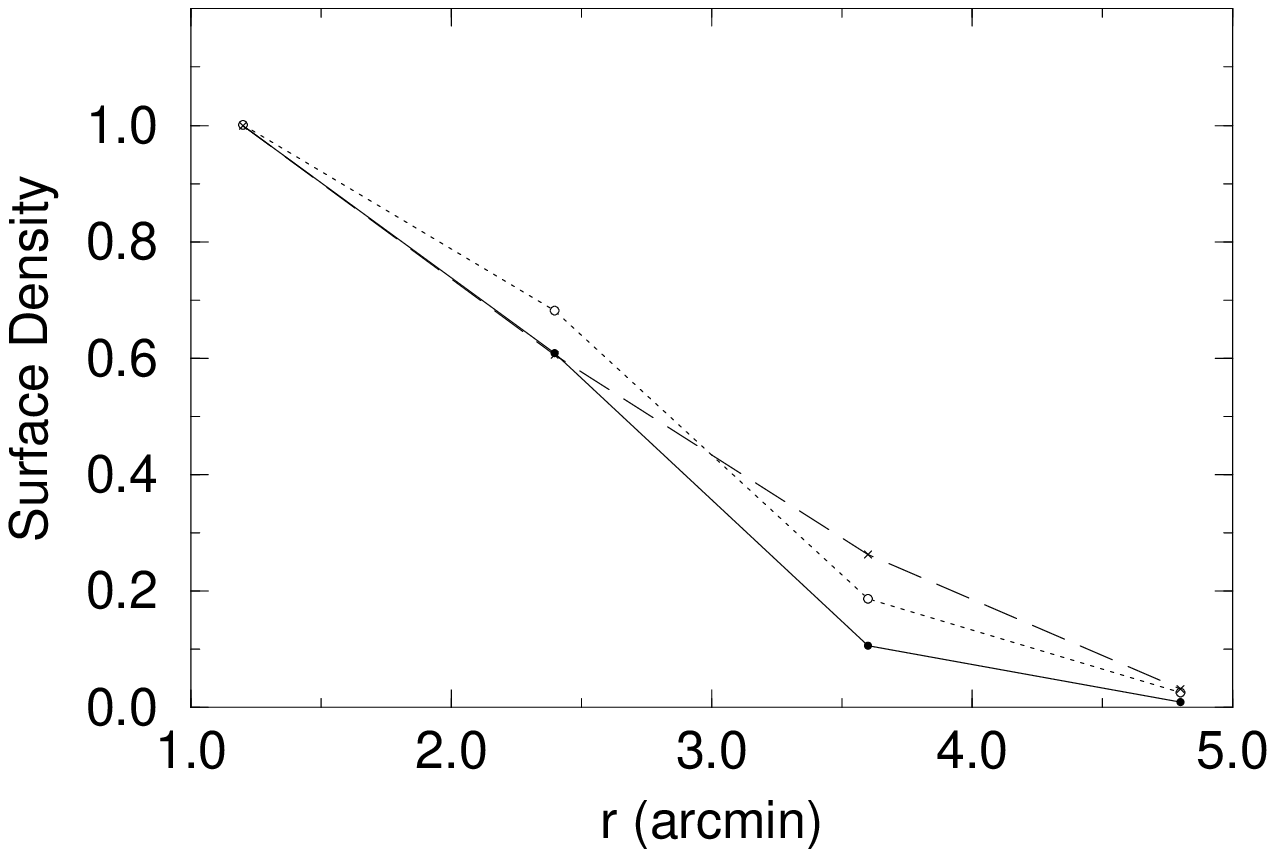}
    \end{center}
     \caption[]{Surface density distributions for three magnitude ranges,
     filled circles (15 $\ge$ $V$ $>$ 14), open circles (16 $\ge$ $V$ $>$ 15)
    and crosses ( 17 $\ge$ $V$ $>$ 16),all distributions have been 
    normalized to 
   the surface density at $r$= 1.2 arcmin. }
   \end{figure}

\subsection{Separation of field stars from cluster members}

The method described above has been used to study the 
membership of the open cluster NGC 4815   
in the four-dimensional space of $(X, Y, V, B-V)$.  
CO94  investigated the degree of completeness for 
their observations, and found that the sample is almost complete  to
$V=17$ mag.
In order to avoid incompleteness problems, 
we only analyze the stars with this limit, thus      
including 739 stars. 
The determination of those stars 
considered as cluster members has a probabilistic nature. 
In Fig. 4, we show the CMD for cluster members with probability 
$P$(c$\mid$${\vec  x_i}$) $>$ 0.5, 0.8, and 0.9, respectively. 
Isochrones (Alongi et al. 1993) of different 
metal abundance $Z$ and age are overplotted.

In order to consider the influence of the observational errors
($\sigma_{V}$ and $\sigma_{B-V}$) in our analysis, we have generated
several sets of 'observed' data by adding an increment ($\Delta$$V$,
and $\Delta$$(B-V)$) to the real values of each star. Each of these
increments has been randomly chosen from a normal distribution of zero
mean and standard deviation equal to the individual observational error.
We found that the  member probabilities are not significantly changed, 
indicating that the observational errors are not very important in
the present analysis.

As mentioned previously, we have used the Arenou et al. (1992) 
extinction law  in our Galaxy  model. 
In order to consider the influence of the uncertainty due to the absorption 
in our analysis, we have checked our results by using the Arp (1965) 
reddening distribution:  
the absorption increases at about a rate of 1.0 mag kpc$^{-1}$ for the first 2 kpc, 
and after 2 kpc, there is no further light absorption.  This simple    
reddening law has been found to be in agreement with the observations 
along several galactic directions in the galactic plane 
(Arp, 1965; Paczynski et al. (1994), Ng \& Bertelli, 1996; Arp \& Cuffey, 1962).  
In Fig. 5, we show the results from the Arp reddening distribution. 
In Fig. 6. we show the $P$(c$\mid$${\vec x_i}$) from the Arenou 
model (P1) against the $P$(c$\mid$${\vec x_i}$) from the Arp reddening 
distribution (P2). 
>From Fig. 6, we can see that, for large P1 and P2 ($>$ 0.8), the difference 
between P1 and P2 is small, but P1 is systematically larger than P2;     
for small P1 and P2 ($<$ 0.5), the difference between P1 and P2 becomes 
larger. For example, with  $P$(c$\mid$${\vec x_i}$) $>$ 80\%,
we found 378 stars with the Arenou et al. (1992) extinction model, 
and 314 stars with the Arp reddening 
distribution.  
Therefore, we can say that the uncertainty due to the 
absorption in our analysis does not change too much the likely open cluster 
members (P(c$\mid$${\vec x_i}$) $>$ 80\%).  
However, we should point out that,  
to derive more reliable membership probabilities, one should have a good
knowledge
about the extinction law in the direction of the cluster.

To our knowledge, the only investigation  in the literature of the
cluster membership for
NGC 4815 is that of CO94. They 
separate cluster stars from field stars by using a radius of
$4^{\prime}.6$ arcmin.
In Fig. 7, we show the CMD for the cluster members by the criterion of 
CO94 for stars with V $<$ 17 mag.   
It can be seen that the use of only  position to determine the cluster 
membership leads to a significant contamination 
by field stars ($V$ $>$ 16 and $B-V$ $>$ 1.5). 
In a standard cluster analysis, identifying cluster members is a process of 
attrition - first selecting by position (or proper motion), then removing 
photometric non-members, then spectroscopic non-members and so forth. 
Our method, combining position and $( V, B-V)$ into a single step, enlarges 
the statistical "distance" between the field stars and cluster members, thus 
reducing the contamination due to the field. Moreover, our method, using   
a probability classification,  can provide an unbiased estimation of the 
intrinsic         
distribution of the cluster stars.

Another 
database for NGC 4815 in the literature is 
that from Kjeldsen \& Frandsen (1991), who obtained 
$UBV$ CCD observations. In Fig.~8, 
we show the colour-magnitude diagram 
for their observations. As pointed out by  Kjeldsen \& Frandsen (1991), 
we can see that the CM diagram  
contains many 
non-members. 
In any case a MS with a turn-off is still clearly visible. We 
found that the isochrone of $Z$ = 0.008 for an age of 500 Myr 
can basically explain the observations of 
Kjeldsen \& Frandsen (1991), which is consistent with our result 
from CO94 sample.

\section{The spatial distribution and luminosity function of NGC 4815}

In this section, the cluster member probability discussed  above 
is used to investigate the 
spatial distribution and luminosity function of NGC 4815. 
Both CO94  and this investigation  
have shown that NGC 4815 has an age of about 500 - 600 Myr, 
which is similar to the age of the Hyades cluster (625 $\pm$ 50 Myr) derived 
by Lebreton et al. (1997) from Hipparcos observations.
It is very 
interesting to compare our results with that of the Hyades cluster,
which has been widely studied and whose properties 
are well understood.

\subsection{Luminosity function}
It is very important to study the cluster
luminosity function because it can provide information about
both the
initial mass function and the cluster dynamical evolution.
Cluster membership probabilities were used to determine the 
luminosity functions for the cluster. The luminosity function  
$\Phi$($M_{v}$) was obtained as:
\begin{equation}
\Phi(M_{v}) = \frac {\Sigma_{j} P(c|x_{j})}{\Delta M_{v}}
\end{equation} 

Where P(c$\mid$${\vec x_j}$) is the membership probability of star $j$, 
$\Delta$$M_v$ is the bin size. 
Table 1 and Fig.~9 show the resulting luminosity function for NGC 4815. 

Our results show that the luminosity function increases at $V$ = 14.0
mag ($M_{v}$ = -0.2) around the MS turnoff, 
with  a prominent drop at
$V$ = 15 mag ($M_{v}$=0.8) that corresponds to the MS gap.
The presence  of a gap in the distribution of stars near the MS turnoff region
is a common feature of old (M67, Montgomery et al 1993)
and intermediate age (NGC 7789, Roger et al. 1994) open clusters.
The gap is considered as a real feature corresponding  to the
evolutionary behaviour of stars at the end of their core H-burning stage
when they undergo the overall
contraction phase (Carraro et al. 1994).
This gap can also be seen in the CMD for all observed stars 
as well (see Fig.~2).
However the luminosity function derived by CO94
did not show this feature, probably due to the presence of
many field stars that were mis-classified as cluster members.
In Fig.~9 we have overplotted the luminosity function
of the Hyades clusters from Reid (1993b) and NGC 7789 from Roger et al. (1994).

\begin{table}
\caption[]{ The luminosity function of the cluster NGC 4815} 
\begin{flushleft}
\normalsize
\begin{tabular}{ccc} \\ \hline
 $V$  &$M_v$    & $\Phi(M_v)$   \\ 
 13.3 & -0.9 &  0.9   \\
 13.5 & -0.7 &  0.8   \\
 13.7 & -0.5 &  2.5   \\
 13.9 & -0.3 &  6.5   \\
 14.1 & -0.1 &  9.2   \\
 14.3 &  0.1 & 10.1   \\
 14.5 &  0.3 & 11.3   \\
 14.7 &  0.5 & 21.1   \\
 14.9 &  0.7 & 12.1   \\
 15.1 &  0.9 & 14.8   \\
 15.3 &  1.1 & 36.5    \\
 15.5 &  1.3 & 42.3   \\
 15.7 &  1.5 & 36.3   \\
 15.9 &  1.7 & 42.0   \\
 16.1 &  1.9 & 49.9   \\
 16.3 &  2.1 & 54.1    \\
 16.5 &  2.3 & 52.1    \\
 16.7 &  2.5 & 53.2   \\
 16.9 &  2.7 & 67.3  \\
\hline
\end{tabular}
\end{flushleft}
\end{table}

Given the luminosity function derived from our data, we can apply 
mass-luminosity relations to derive an estimate of the total mass of the 
NGC 4815. We have used  the relation 

\begin{equation}
log ({\cal M}) = -0.067 M_v + 0.356  ~~~~~~~~~~~~~~ M_v \le 0.22
\end{equation}

\begin{equation}
log ({\cal M}) = -0.163 M_v + 0.558   ~~~~~~~~~~~~~~ M_v > 0.22
\end{equation}

derived from Andersen's data (1991).  
>From masses ({$\cal M$}, in solar mass)  
and membership probabilities, the observed mass in the cluster 
can be determined from $\Sigma_j$ P(c$\mid$${\vec x_j}$){$\cal M_j$}. 
To the limit of the photometry, we find a mass of 
the cluster of 912 solar masses. In Fig. 10, we show the derived 
mass function ($N (log({\cal M}))$) for the NGC 4815. 
By using a least-squares fit, we found that the mass function has a slope of 
-2.2 $\pm$ 0.3 for 1.5 $<$ ${\cal M}$ $<$ 2.3 in the plane of 
($log$ ($N$), $log$(${\cal M}$). Francic (1989) has derived the slope for the 
composite mass function for six clusters with a mean age of 400 Myr, he found d $log$ $N$/d $log$ 
(${\cal M}$)  
= -1.97 $\pm$ 0.17 for 1.1 $<$ ${\cal M}$ $<$ 2.5. His result is in agreement 
with our result in this paper,  considering the error bars.   
Our result is also in good agreement with the value determined by 
Scalo (1986) for the field stars (-2.0 $\pm$ 0.18 for 0.8 $<$ ${\cal M}$ $<$ 
18).

\subsection{The spatial distribution of cluster members}
We have taken annuli with a width
of 0.3 arcmin centered on the cluster center and have counted the numbers
of stars ($N$ = $\Sigma_{j}$ $P$(c$\mid$${\vec x_j}$) falling within each annulus.
Fig.~11
shows  the radial density distribution for the cluster NGC 4815,
where the error bars were calculated assuming that the
number of stars in a bin is governed by Poisson statistics.

It was shown by Van den Bergh \& Sher (1960) that the projected surface
distribution of stars for an open cluster can be represented by
$\rho (r) = \rho_{0} exp (-r/r_{0})$, where $r_{0}$ is the
scale length for the projected surface density of cluster members.
We have used this exponentially decreasing function of the distance from 
the cluster center to fit our results, and found the scale length 
to be $r_{0}$ = 1.67 $\pm$ 0.06 arcmin. Since the distance to NGC 4815 
is known, we can transform the angular scalelength to a linear distance. 
Assuming that the cluster is symmetric, we can calculate the average volume 
density of stars within the cluster. For the inner region of the 
cluster ($r$ $<$ 1.67 arcmin), we derive an average density of   
47.5 stars/pc$^3$ and a mean separation of 0.171 pc. If we calculate 
the density at the edge of the cluster, taken to be at a diatance 
of 4.0 arcmin (see Fig. 11), we find an average density of 7.9 stars/pc$^3$ 
with a mean separation of 0.312 pc.

Reid (1992) has studied the Hyades cluster members and found 
that the brighter stars are more centrally concentrated 
(mass segregation).  
In order to look for similar evidences in NGC 4815, 
we have 
divided our cluster members into three magnitude ranges, 
15 $\ge$ $V$ $>$ 14; 16 $\ge$ $V$ $>$ 15; 17 $\ge$ $V$ $>$ 16. 
In Fig.~12 we  plot the surface density distributions for these 
three magnitude ranges, all distributions have been normalized to the 
surface density at $r$= 1.2 arcmin. From Fig. 12, we 
can see that the brighter stars are slightly more  
centrally concentrated, as one would expect from mass segregation.
However, because the sample used here is limited with $V$ $\le$ 17 mag, the
observational evidence  is limited. Deeper samples are needed to confirm
this result.

\section{Conclusions}

We have developed a new algorithm, which uses  
both positional and photometric data simultaneously to search for
open cluster members. The algorithm  
can be extended to use all possible variables, including proper
motions, 
radial velocity and multi-color photometry automatically.  
When the contamination of field stars is not very serious, 
then simple methods can be used to identify cluster members 
(radial distance from the centre; position in the 
colour-magnitude diagram; etc); if field star contamination 
is significant, then  one requires a more sophisticated technique.   
The method developed in this work, which uses all variables simultaneously 
to search for open cluster members, gives better results. This is 
particularly true when 
cluster stars lie beyond the observed sample that is used, where the
 contamination of field stars is difficult to  estimate.

The algorithm  has been used to isolate  
open cluster members of NGC 4815 from the field stars. We have compared 
our results with those of CO94. 
A clean and well defined colour-magnitude diagram (CMD) for the 
cluster has been obtained from our new algorithm. 
After determining the cluster member probability, for the first time we have 
investigated the spatial
distribution of the
cluster NGC 4815 and its luminosity function. To summarize, we find that:
\begin{enumerate}
\item NGC 4815 is a cluster of the Hyades generation, with an age of about 
500 - 600 Myr.
\item the projected  surface distribution of stars is
well-represented by an
exponentially decreasing function 
with a scale length of 1.67 $\pm$ 0.06 arcmin.
\item there is evidence for mass segregation.
\item 
the luminosity function for NGC 4815 is similar to that of Hyades cluster, 
and shows a gap at $V$ = 15 mag. 
\item to the limit of the  photometry, the observed mass is of about 
900 solar masses for 
NGC 4815 
\item the mass function has a slope of -2.2 $\pm$ 0.3 for 
1.5 $<$ {$\cal M$} $<$ 2.3 in the plane of ($log$ ($N$), $log$ ({$\cal M$})).  
\end{enumerate}

These observational results are important, and can be used to
constrain models of stellar evolution (Carraro et al. 1994).
Other contaminated open clusters are going to be analyzed in the future.

\begin{acknowledgements}
This research was started when one of us (BC) was invited to visit 
Beijing Astronomical Observatory (BAO) by the National 
Natural Science Foundation of China.   
We thank Drs. S. Ortolani, F. Figueras and L.C. Deng for helpful comments. 
BC acknowledges financial support from the Ministerio de Educati\'on y 
Ciencia. 
This work has been supported by CICYT under contract PB95-0185.
\end{acknowledgements}

{}
\end{document}